\documentclass[intlimits,twoside,a4paper]{article}

\usepackage[cp1251]{inputenc}

\usepackage[eqsecnum]{cmpj3}

\usepackage{bm}

\issue{2019}{22}{4}{43603}
\doinumber{10.5488/CMP.22.43603}

\title[IXS from salol]%
{Inelastic x-ray scattering reveals the ergodic to nonergodic transition of salol, a liquid with local order}
\author[L. Comez \textsl{et al.}]{L. Comez\refaddr{label1},
        D. Fioretto\refaddr{label2}, J. Gapinski\refaddr{label3}, G. Monaco\refaddr{label4}, A. Patkowski\refaddr{label3}, W. Steffen\refaddr{label5}}
\addresses{
\addr{label1} IOM-CNR c/o Dipartimento di Fisica e Geologia, Universit\'a di Perugia, 06123 Perugia, Italy 
\addr{label2} Dipartimento di Fisica e Geologia, Universit\'a di Perugia, 06123 Perugia, Italy
\addr{label3} Faculty of Physics, A. Mickiewicz University, Uniwersytetu Pozna\'nskiego 2, 61-614 Pozna\'n, Poland 
\addr{label4} Department of Physics, University of Trento, Via Sommarive 14, 38123 Povo, Italy
\addr{label5} Max Planck Institute for Polymer Research, P.O. Box 3148, 55128 Mainz, Germany
}

\date{Received June 25, 2019}

\begin{document}

\maketitle

\begin{abstract}
We have studied the high-frequency dynamics of salol by inelastic x-ray scattering over a wide temperature range between 50 and 450 K, across the glass transition. We find that salol efficiently realizes the mechanism of dynamical arrest described by the mode-coupling theory, as manifested by a cusp singularity in the behaviour of the non-ergodicity parameter and a $Q$ dependence of the critical non-ergodicity parameter that is in phase with the static structure factor. These results confront positively the mode-coupling theory with liquids with local order.
\pacs 64.70.Pf, 78.70.Ck, 61.20.Lc
\end{abstract}

\section{Introduction}

At the turn of the second and third millennium, a great deal of work, both theoretical and experimental, has been devoted to the study of the physics of disordered systems with particular emphasis on the processes connected with the structural arrest and glass transition.

The mode coupling theory (MCT) was introduced to provide a self-consistent treatment of the structural arrest in simple liquids \cite{Goetze99}. The idealized version of MCT describes the glass transition as an ergodic to non-ergodic transition occurring at a critical temperature $T_\text{c}$, associated with a singular behaviour of the long-time limit of the normalized density correlator, the so-called non-ergodicity factor~$f_Q$. The peculiar temperature ($T$) and wave vector ($Q$) dependences expected for $f_Q$ are: i) a square-root temperature behaviour below $T_\text{c}$, $f_Q(T)=f_Q^\text{c}+h_Q\sqrt{(T-T_\text{c})/T_\text{c}}$, where $f_Q^\text{c}$ is the critical non-ergodicity parameter and $h_Q$ is the critical amplitude at a given wavevector $Q$; ii) a $Q$ dependence of $f_Q^\text{c}$ and of $h_Q$, which are in phase and in antiphase with the static structure factor $S(Q)$, respectively. 

In the same years, the inelastic x-ray scattering (IXS) technique was developed, capable of detecting the dynamic structure factor $S(Q,\omega)$ of glasses \cite{Benassi96} and glass forming liquids in the nm$^{-1}$ $Q$-range~\cite{Sette98}. IXS, together with neutron scattering and MD simulations, became the method of choice for the quantitative test of MCT predictions. 
Most of the tests were performed in simple liquids, such as van der Waals molecular liquids, with relatively simple relaxation patterns \cite{Monaco98}. More complex systems were also analyzed, such as polymers \cite{fioretto99,Frick90}, and a good coherence with MCT was found when the contribution of the structural relaxation was singled out with respect to those of secondary processes \cite{Fioretto02}, complementing IXS with Brillouin light scattering measurements \cite{Comez12}. An interesting class of liquids is that of associated liquids, where local order extends over several neighboring molecules, giving rise to a pre-peak in the low-$Q$ region of $S(Q)$. A pioneering work performed by some of us on the associated liquid m-toluidine provided experimental evidence that hydrogen bond clustering can coexist with the signature of the ergodic to non-ergodic transition predicted by the MCT \cite{Comez05,Comez06}.  

In the present work, we extend this investigation to salol, a widely studied glass-forming system~\cite{Comez04,Comez02,Kalam03,Kalam03b}.

\section{Experiment}

Salol (Phenyl salicylate, Fluka, purity $>98$\%) was dried under vacuum for three days at $95^{\circ}$C.  To remove dust as potential sites of heterogeneous nucleation, the sample was then filtered through a 0.22~{\textmu}m Durapore (Millipore company) filter into dust free vials which were sealed until used in the actual experiment to fill the sample cells.

IXS experiments were performed at the very high energy resolution beamlines ID16 and ID28 of the European Synchrotron Radiation Facility (ESRF), Grenoble. The monochromator and analyzer crystals were operated at backscattering configuration corresponding to an incident photon energy of 21.747 keV and a total energy resolution of 1.5 meV. Spectra were taken at $Q$-values between 1 and 15 nm$^{-1}$ in steps of 1 nm$^{-1}$ at temperatures between 50 and 450~K.  All spectra were corrected for scattering of the empty cell and normalized to the monitored incoming intensity. 

\begin{figure}[!b]
\vspace{-3mm}
\centerline{\includegraphics[width=0.78\textwidth]{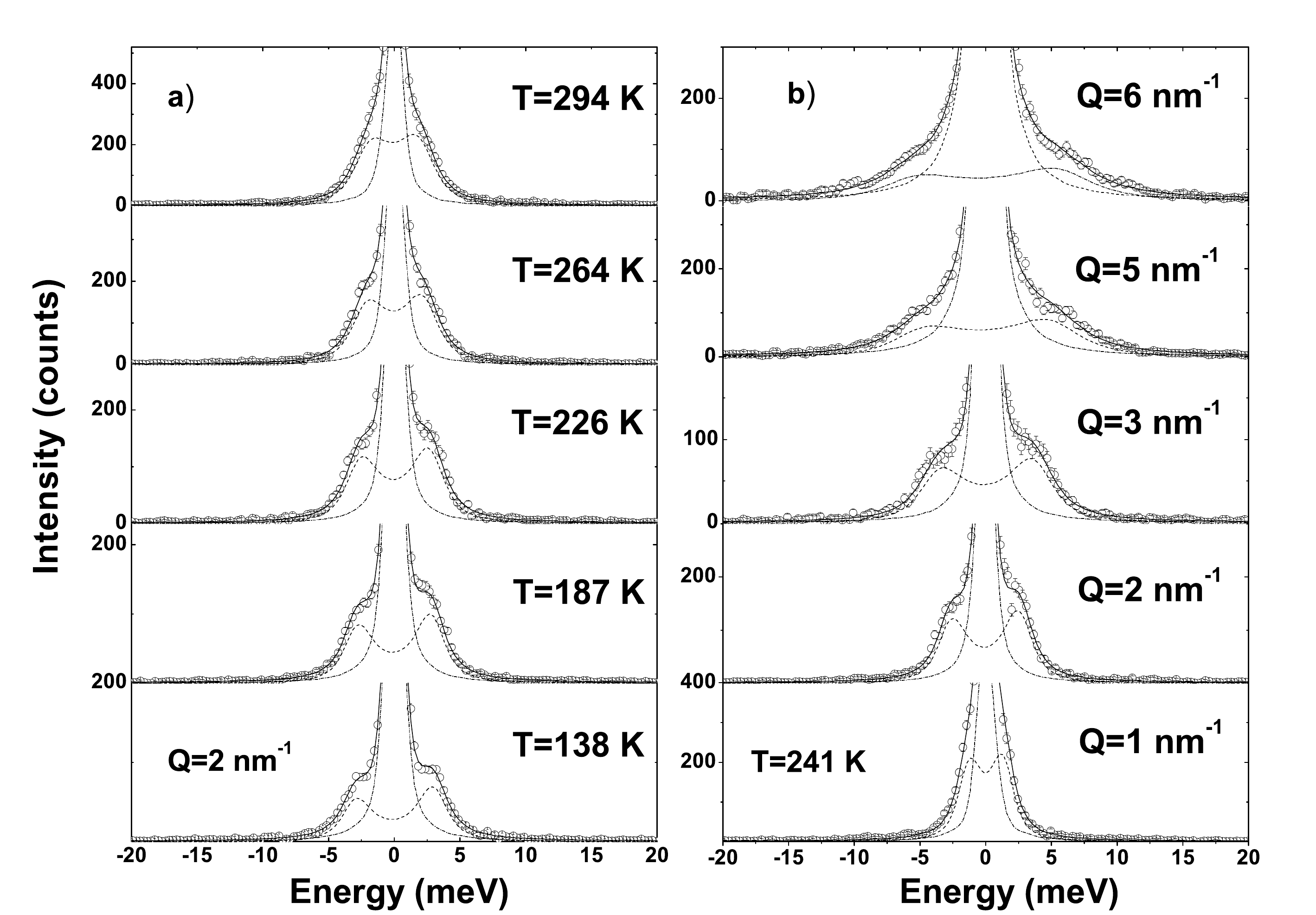}}
\caption{Left-hand panel: IXS spectra of salol taken at $Q= 2$ nm$^{-1}$, at the indicated temperatures. Right-hand panel: IXS spectra of salol taken at $T = 241$ K, at the indicated $Q$ values. The fitting curves (solid lines), the quasi-elastic contributions (dash-dotted lines) and the inelastic contributions (dashed lines) to the total fit are reported together with the data points (open circles).} \label{fig1}
\end{figure}

In figure~\ref{fig1} we report some IXS spectra for selected temperatures at the fixed exchanged wave vector $Q = 2$ nm$^{-1}$ (left-hand panel), and some spectra obtained for different values of $Q$ at $T = 241$ K (right-hand panel).

\section{Results and discussion}

All measured spectra show quasielastic and inelastic contributions, whose characteristic parameters were obtained by fitting the convolution of the instrumental resolution function $R(\omega)$ with a model for $S(Q,\omega)$ including a delta function for the quasielastic line and a damped harmonic oscillator for the two inelastic side peaks \cite{Corezzi06}, which are due to the Brillouin scattering of photons by longitudinal acoustic (LA) modes propagating in salol. From this fitting procedure, the characteristic frequency $(\Omega)$ and linewidth $(\Gamma)$ (FWHM) of the LA modes were obtained, together with the intensities of the quasielastic and inelastic contributions, $I_\text{el}(Q)$ and $I(Q)$.

\begin{figure}[!t]
	\centerline{\includegraphics[width=0.45\textwidth]{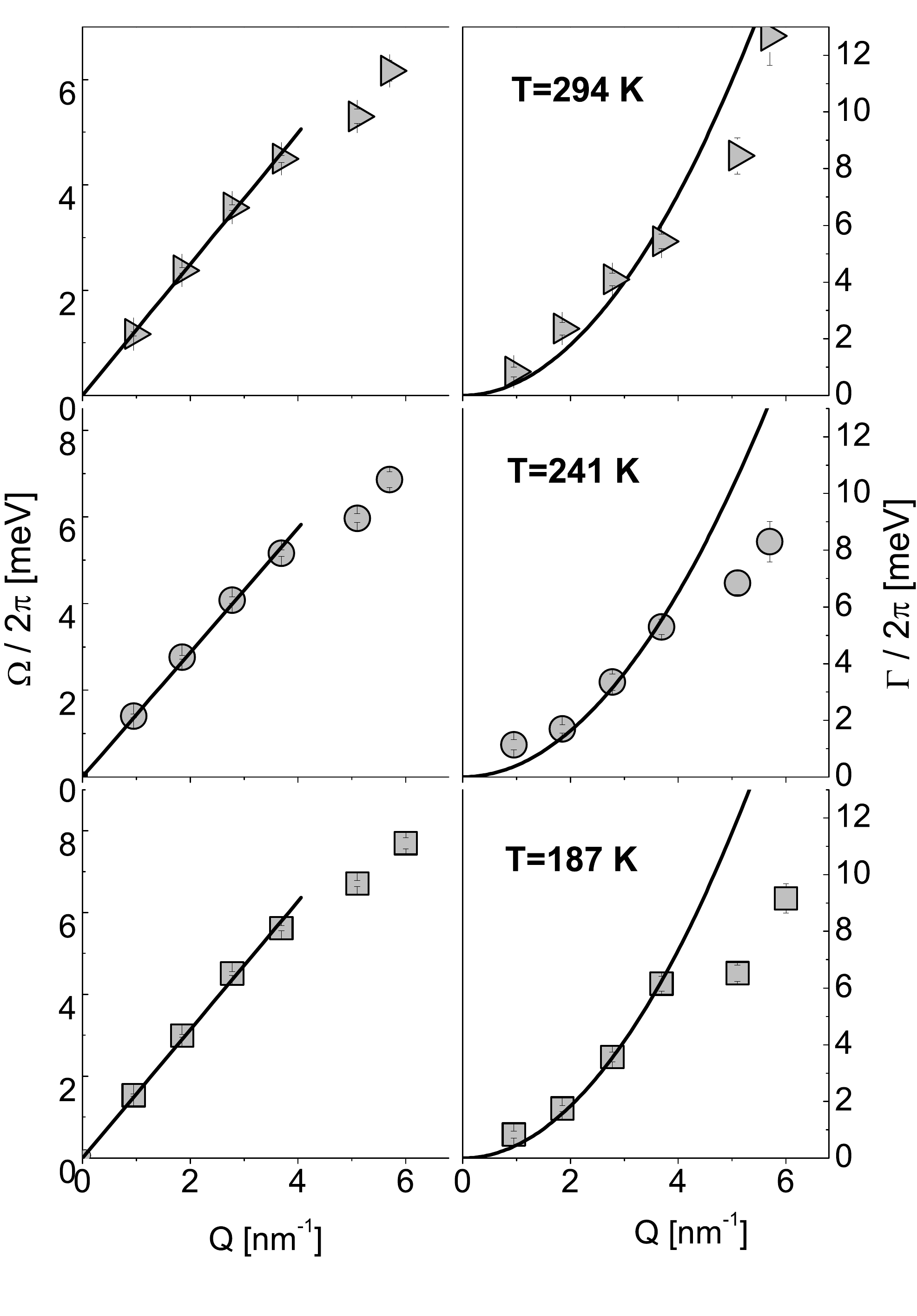}}
	\caption{Left-hand panels: Linear dependence on $Q$ of the longitudinal acoustic frequency in the low-$Q$ range. Right-hand panels: In the $Q$ region, where the acoustic dispersion relation is linear, the broadening of the acoustic excitations shows a $T$-independent $Q^2$ behaviour.} \label{fig2}
\end{figure}

Figure~\ref{fig2} shows the values of $\Omega$ and $\Gamma$ obtained by the fitting procedure, as a function of $Q$ and for three selected temperatures. The almost linear behaviour of $\Omega$ vs. $Q$ allows us to estimate the average value for the velocity $(v)$ of the LA modes in the low $Q$ regime by fitting to the first four data points the expression:  $\Omega =v \cdot Q$. In the same region, the almost quadratic behaviour of $\Gamma$ vs. $Q$ gives an estimate of the unrelaxed value of the longitudinal kinematic viscosity $D_\text{L}$ and thermal diffusion $D_\text{T}$ through $D_\text{L}+(\gamma-1)D_\text{T} = \Gamma/Q^2 $ with $\gamma$ being the ratio of specific heats.

\begin{figure}[!t]
	\centerline{\includegraphics[width=.98\textwidth]{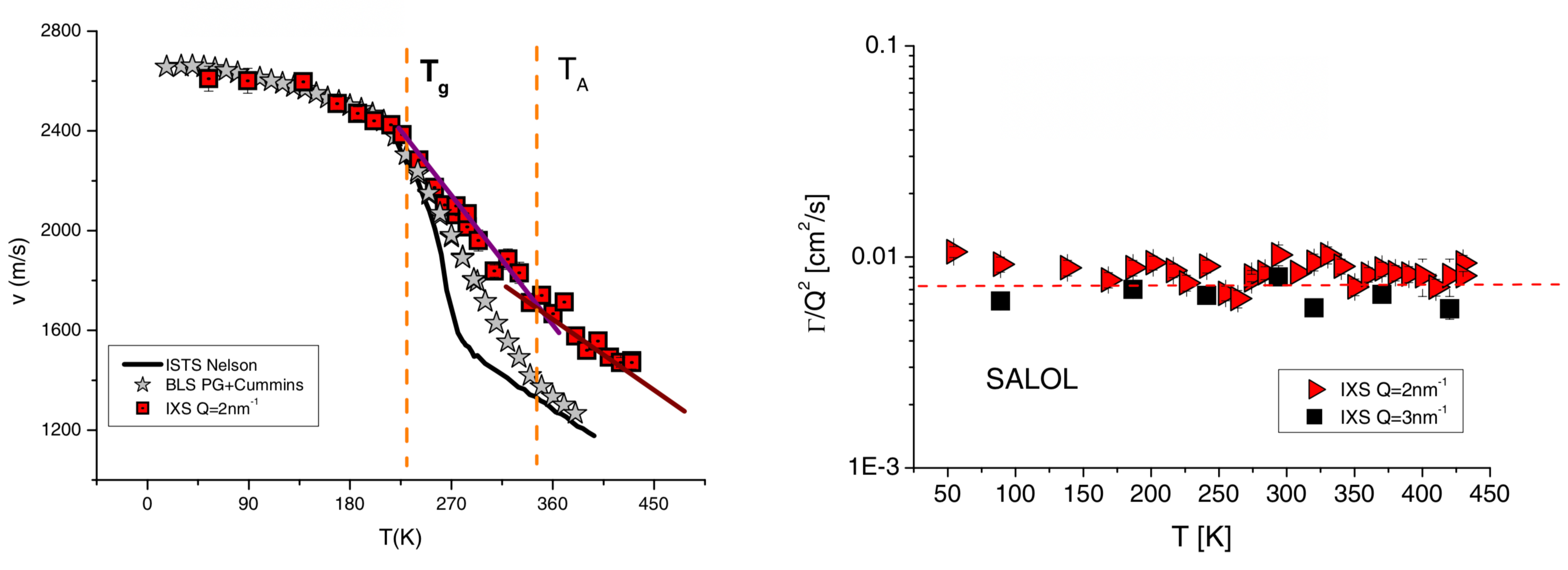}}
	\caption{(Colour online) Left-hand panel: Limiting high frequency longitudinal sound velocity (squares) determined by a DHO~$+$~delta function analysis of IXS spectra compared with the literature data. Evidence is given for two changes in the slope corresponding to $T_\text{g}$ and $T_\text{A}$. Right-hand panel: Limiting high frequency longitudinal kinematic viscosity.} \label{fig3}
\end{figure}

Velocities and viscosities obtained by this method are reported in figure~\ref{fig3}. 

In the low temperature regime, the velocity of the LA modes obtained by IXS agrees very well with the unrelaxed sound velocities $(v_{\infty})$ measured at much lower $Q$ values by Brillouin light scattering (BLS)~\cite{Bencivenga,Zhang04} and impulsive stimulated thermal scattering (ISTS) techniques \cite{ISTS}. 
For increasing temperature, a change of the slope of $v(T)$ is visible close to the glass transition of salol $T_\text{g} = 220$~K, where a more pronounced temperature dispersion starts to occur \cite{Maciovecchio98}. 
Interestingly, the temperature behaviour of $v_\text{IXS}$ is also compatible [see linear extrapolations in figure~\ref{fig3}~(a)] with the existence of a further transition in the liquid at $T_\text{A} = 348$ K. $T_\text{A}$ was proposed and shown as the transition temperature of the $\alpha$-process from VFT to Arrhenius behaviour \cite{Hansen98,Eckstein00}.

At the highest investigated temperatures, the values of $v_\text{BLS}$ and of $v_\text{ISTS}$ progressively approach those of the relaxed sound velocity $v_0$. Conversely, the values of $v_\text{IXS}$ remain considerably higher than $v_0$, suggesting that IXS probes the unrelaxed sound velocity $v_{\infty}$ also in the liquid phase. This is also supported by the temperature dependence of the linewidth of Brillouin peaks reported in figure~\ref{fig3}~(b). In fact, it can be seen that $\Gamma/Q^2$ at the two lowest probed $Q$ values is almost constant in the whole investigated temperature range suggesting that the mechanism responsible for the peak broadening in the glassy phase continues to dominate also at temperatures higher than  $T_\text{g}$. This result can be explained by the fact that, at the probed $Q$s, the phonon frequency is much higher than the rate of the structural relaxation (unrelaxed condition) and, therefore, the sound waves probe the system as ``frozen'' in the whole investigated temperature range. In this regime, the broadening of the Brillouin peaks is due to the disordered molecular structure rather than due to truly dynamic processes \cite{Ruocco99}. 

\begin{figure}[!b]
	\centerline{\includegraphics[width=0.68\textwidth]{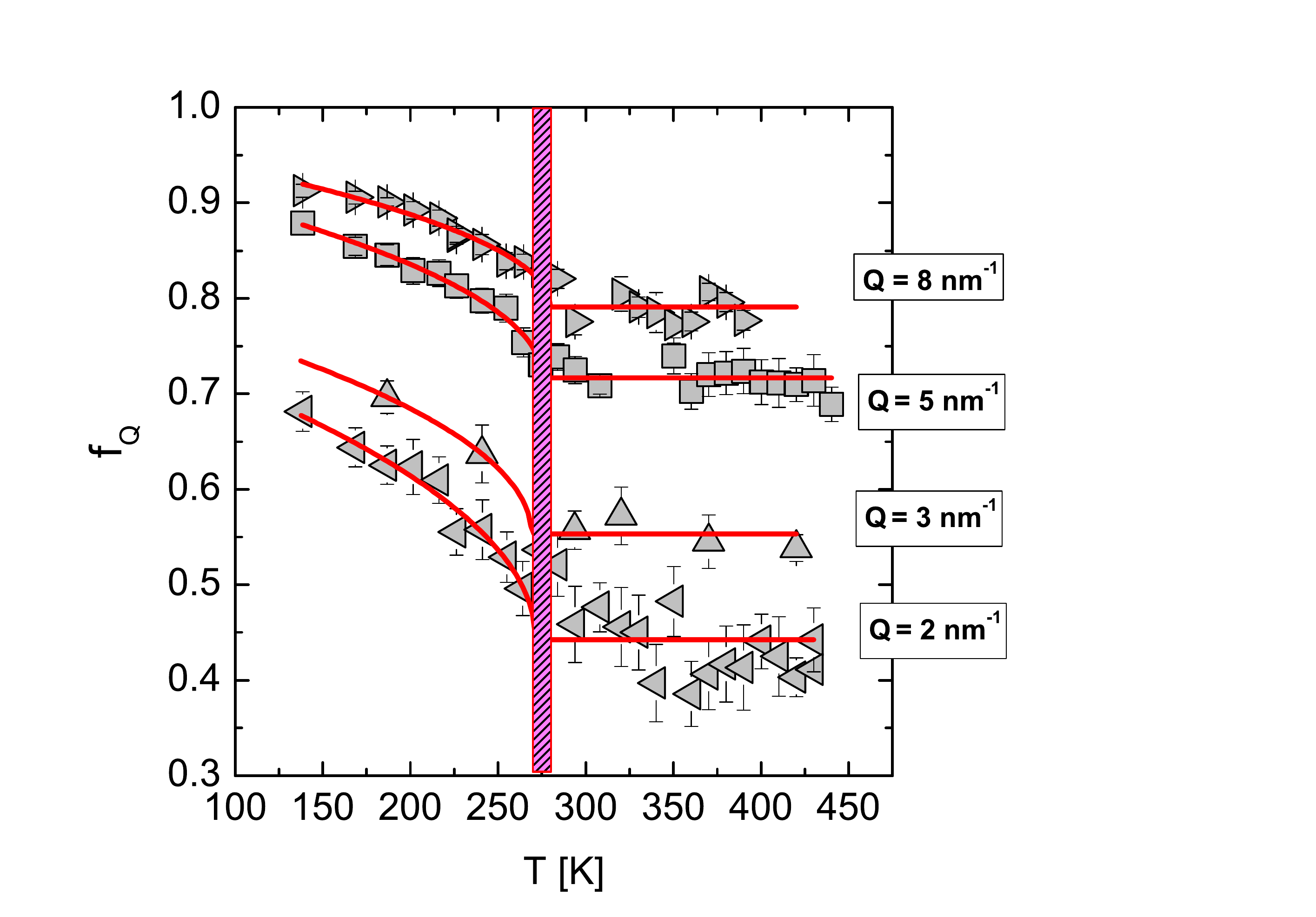}}
	\caption{(Colour online) Temperature dependence of the effective non-ergodicity factor $f_Q$ of salol for selected values of the exchanged wavevector $Q$. The red line is a guide for eyes to indicate the critical temperature $T_\text{c} = 273 \pm 5$ K. The solid lines are the best fits obtained using the square-root function predicted by the mode coupling theory \cite{Goetze99}.} \label{fig4}
\end{figure}

An important consequence of the unrelaxed regime here probed by IXS is that the contribution of the structural relaxation to $S(Q,\Omega)$ is all included within the area of the quasielastic peak. In this condition, the relative amplitude of the structural relaxation, i.e., the non-ergodicity parameter $f_Q$, can be obtained from the IXS spectra as the ratio of the intensities $f_Q = I_\text{el}(Q) / [I_\text{el}(Q) + I(Q)]$, and the analysis of the IXS spectra becomes a powerful tool to test the predictions of MCT.

Figure~\ref{fig4} reports the temperature behaviour of $f_Q$ at four different values of $Q$, showing clear evidence of the square-root singularity predicted by the idealized MCT \cite{Goetze99}, even more clearly than in the previously reported m-toluidine case \cite{Comez05}, due to the improved signal-to-noise ratio of the present IXS spectra.  It is worth noting that the value of the critical temperature $T_\text{c} = 273 \pm 5$ K is $Q$ independent within experimental error, and that its value favourably agrees with previous estimates obtained from dielectric spectroscopy \cite{Stickel95} and from wide angle x-ray experiments combined with MD simulations \cite{Eckstein00}.

\begin{figure}[!t]
	\centerline{\includegraphics[width=0.55\textwidth]{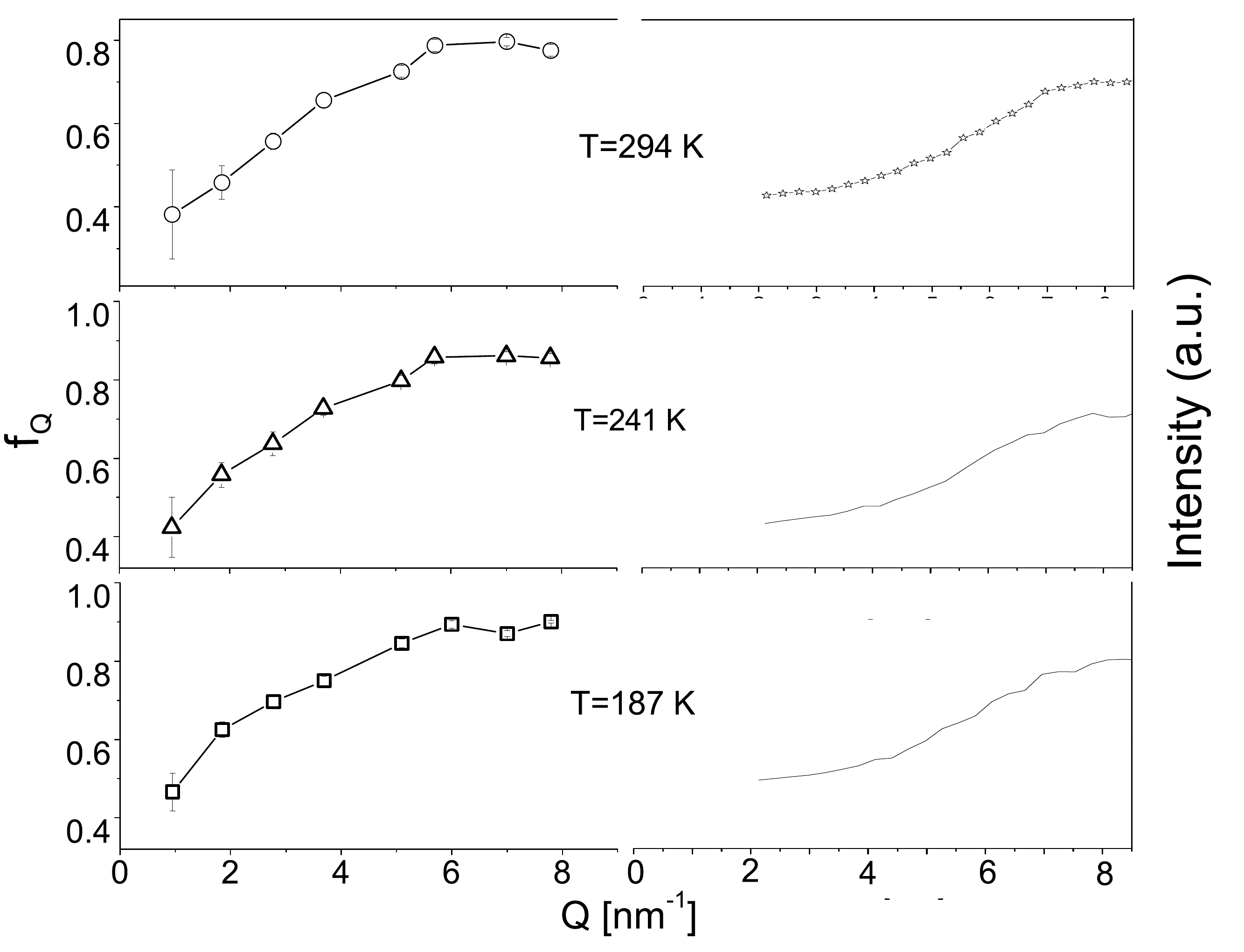}}
	\caption{Left-hand panel: The non-ergodicity parameter $f_Q$ of salol at different temperatures. Right-hand panel: $I(Q)$ obtained by wide angle x-ray scattering measurements. } \label{fig5}
\end{figure}

\begin{figure}[!b]
	\centerline{\includegraphics[width=0.65\textwidth]{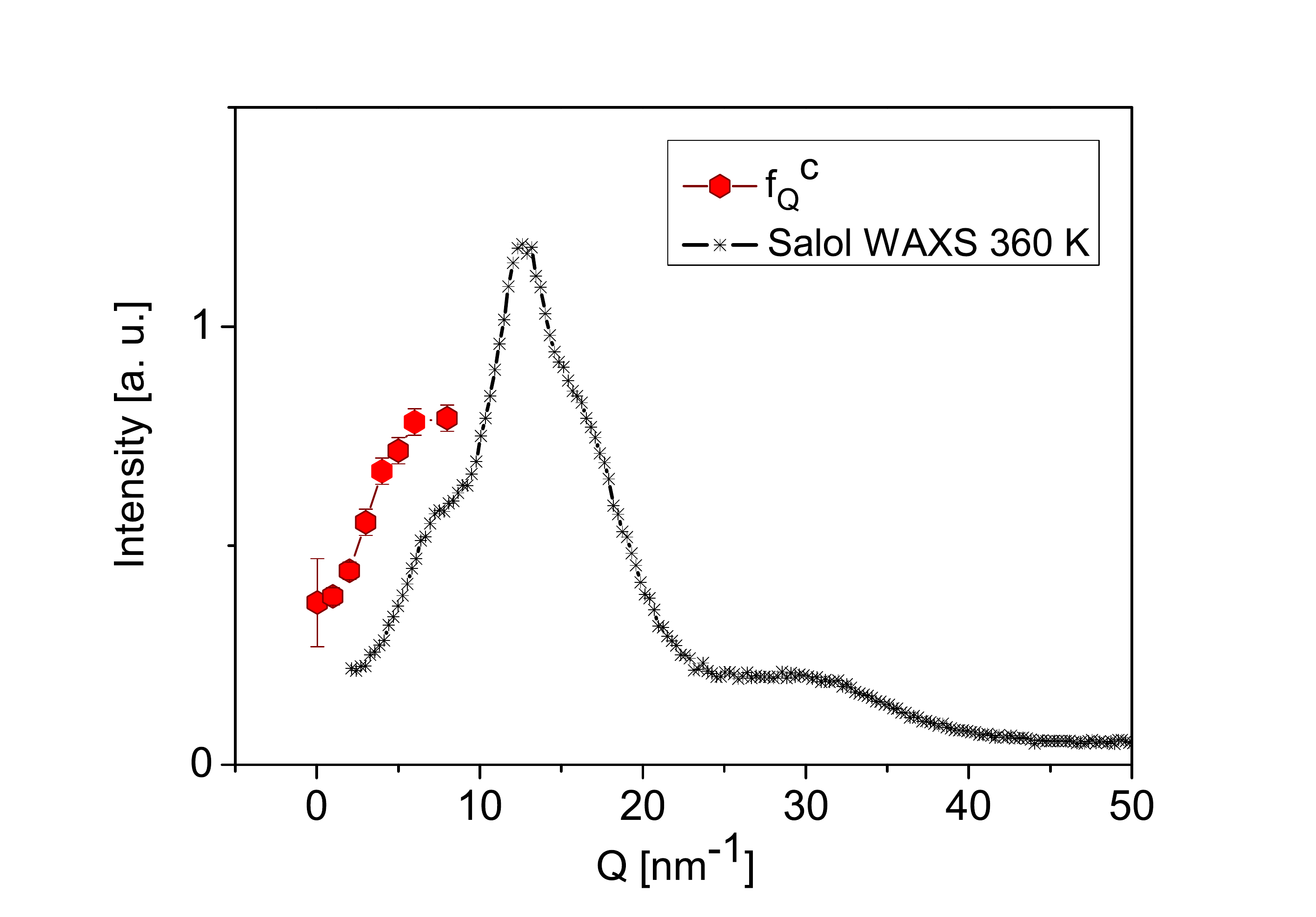}}
	\caption{(Colour online) The plateau of the non-ergodicity factor $f^\text{c}_Q$ oscillates in phase with the static structure factor. } \label{fig6}
\end{figure}

Figure~\ref{fig5} shows that $f_Q$ changes in phase with the static structure factor, in agreement with previous observations \cite{Comez05}. 
For $T > T_\text{c}$, MCT predicts a plateau in the temperature dependence of $f_Q$, i.e., $f_Q(T)= f_Q^\text{c}$, which is also clearly visible in figure~\ref{fig4}.  The values of $f_Q^\text{c}$, on the whole, follow in phase the oscillations of the static structure factor (figure~\ref{fig6}), coherent with MCT calculations for simple liquids.

Based on these two evidences, consistent with the results previously obtained in m-toluidine, we can infer that the predictions of MCT for the existence of a square-root singularity at $T_\text{c}$ and for its $Q$ dependence are robust enough to be also observed in clustering systems \cite{Baran10,Eckstein00}.

A peculiar property of salol and m-toluidine which probably makes the cusp behaviour of the non-ergodicity factor more visible than in other associated liquids is their high degree of fragility. In fact, a phenomenological analysis of IXS spectra of glass-forming systems has revealed that the stronger is a glass former, the higher is the non-ergodicity factor around the glass transition \cite{Scopigno03,Buchenau04}. The non-ergodicity factor can be also expressed as the relative longitudinal modulus variation at $T_\text{g}$ \cite{Buchenau04}. For a strong glass former, like SiO$_2$, the value of $f_Q$ at $T_\text{g}$ is close to unity, and thus the amplitude of the square-root cusp is hardly visible within experimental error. Conversely, for fragile glass-formers such as m-toluidine and salol, fragility favours the visibility of temperature and $Q$-dependences of the non-ergodicity parameter, as shown here.

\section{Conclusions}
In conclusion, our findings corroborate the notion that the early stage of the structural arrest, where the cage effect dominates the molecular dynamics of simple liquids, shows a universal character --- the square-root cusp of the non-ergodicity factor at a critical temperature $T_\text{c}$ --- which is also shared by liquids with a local order \cite{Comez05}. This is a non-trivial result since it suggests that the structural arrest which occurs in dense, simple liquids, shares some common, if not “universal”, features with the structural arrest of locally ordered liquids, where the coordination number is determined by local symmetry constraints and where the mere significance of the cage effect is questionable. To this respect, it is interesting to recall that an IXS investigation of even more locally-structured glass formers, namely reactive binary mixtures, has shown the same signatures of the ergodic to non-ergodic transition predicted by the MCT \cite{Corezzi06}. In that system, in analogy to what is found in supercooled simple liquids, the ergodicity breakdown is observed far before the experimental glass transition, and the static structure evolving during the covalent bond formation is found to consistently affect the wave vector dependence of the non-ergodicity factor. It remains a current challenge to identify the appropriate MCT model that describes the structural arrest in presence of molecular association or even induced by chemical bonds.

\section*{Acknowledgements}

Special thanks to Giancarlo Ruocco, who largely inspired our research on glasses.

\ukrainianpart

\title{Непружне рентгенівське розсіювання розкриває перехід з ергодичного у неергодичний стан в салолі,
рідині з локальним впорядкуванням}
\author{Л. Гомез\refaddr{label1}, Д. Фіоретто\refaddr{label2}, Й. Гапінскі\refaddr{label3}, Дж. Монако\refaddr{label4}, A. Патковскі\refaddr{label3}, В. Штеффен\refaddr{label5}}
\addresses{
\addr{label1} IOM-CNR при Факультеті фізики і геології, Університет Перуджі, 06123 Перуджа, Італія
\addr{label2} Факультет фізики і геології, Університет Перуджі, 06123 Перуджа, Італія
\addr{label3} Факультет фізики, Університет A. Miцкевича, 61-614 Познань, Польща
\addr{label4} Фізичний факультет, Університет Тренто, 38123 Пово, Італія
\addr{label5} Інститут Макса Планка для полімерних досліджень, п/с 3148, 55128 Майнц, Німеччина
}

\makeukrtitle

\begin{abstract}

Ми дослідили високочастотну динаміку салола непружним розсіюванням рентгенівських променів в широкій області 
температур між 50 і 450 K, що покриває перехід в стан скла. Ми знайшли, що салол ефективно реалізує механізм
динамічного арешту, який описується теорією взаємодіючих мод, що проявляється сингулярністю в поведінці параметру
неергодичності та в $Q$-залежності критичного параметру неергодичності, що узгоджується зі статичним структурним
фактором. Ці результати позитивно співставляють теорію взаємодіючих мод з рідинами, які мають локальне впорядкування.

\keywords  перехід в стан скла, розсіювання рентгенівських променів, теорія взаємодіючих мод 

\end{abstract}


\begin{thebibliography}{99}
\bibitem{Goetze99} Gotze W., Sjogren L., Rep. Prog. Phys., 1992, \textbf{55}, 241--376, \doi{10.1088/0034-4885/55/3/001}.
\bibitem{Benassi96}	Benassi P., Krisch M., Masciovecchio C., Mazzacurati V., Monaco G., Ruocco G., Sette F., Verbeni R.,  Phys. Rev. Lett., 1996, \textbf{77}, 3835--3838, \doi{10.1103/PhysRevLett.77.3835}. 
\bibitem{Sette98}	Sette F., Krisch M.H., Masciovecchio C., Ruocco G., Monaco G., Science, 1998, \textbf{280}, 1550--1555,  \doi{10.1126/science.280.5369.1550}.
\bibitem{Monaco98}	Monaco G., Masciovecchio C., Ruocco G., Sette F., Phys. Rev. Lett., 1998, \textbf{80}, 2161--2164,\\ \doi{10.1103/PhysRevLett.80.2161}.
\bibitem{fioretto99} Fioretto D., Buchenau U., Comez L., Sokolov A., Masciovecchio C., Mermet A., Ruocco G., Sette F., Willner~L., Frick B., Richter D.,  Verdini L., Phys. Rev. E, 1999, \textbf{59},  4470--4475, \doi{10.1103/PhysRevE.59.4470}.  
\bibitem{Frick90}	Frick B., Farago B., Richter D., Phys. Rev. Lett., 1990, \textbf{64}, 2921--2924,
\doi{10.1103/PhysRevLett.64.2921}.
\bibitem{Fioretto02}	Fioretto D., Mattarelli M., Masciovecchio C., Monaco G., Ruocco G., Sette F., Phys. Rev. B, 2002, \textbf{65}, 224205--224210, \doi{10.1103/PhysRevB.65.224205}.
\bibitem{Comez12}	Comez L., Masciovecchio C., Monaco G., Fioretto D., Solid State Phys., 2012, \textbf{63}, 1--77,\\ \doi{10.1016/B978-0-12-397028-2.00001-1}.
\bibitem{Comez05}	Comez L., Corezzi S., Monaco G., Verbeni R., Fioretto D., Phys. Rev. Lett., 2005, \textbf{94}, 155702, \doi{10.1103/PhysRevLett.94.155702}.
\bibitem{Comez06}	Comez L., Corezzi S., Monaco G., Verbeni R., Fioretto D., J. Non-Cryst. Solids, 2006, \textbf{352}, 4531--4535, \doi{10.1016/j.jnoncrysol.2006.01.164}.
\bibitem{Comez04}	Comez L., Corezzi S., Fioretto D., Kriegs H., Best A., Steffen W., Phys. Rev. E, 2004, \textbf{70}, 011504, \doi{10.1103/PhysRevE.70.011504}.
\bibitem{Comez02}	Comez L., Fioretto D., Kriegs H., Steffen W.,  Phys. Rev. E, 2002, \textbf{66}, 032501, \doi{10.1103/PhysRevE.66.032501}.
\bibitem{Kalam03} Kalampounias A.G., Kirillov S.A., Steffen W., Yannopoulos S.N., J. Mol. Struct., 2003, \textbf{651}, 475--483, \doi{10.1016/S0022-2860(03)00128-5}.
\bibitem{Kalam03b} Kalampounias A., Yannopoulos S.N., Steffen W., Kirillova L.I., Kirillov S.A., J. Chem. Phys., 2003,  \textbf{118}, 8340--8349, \doi{10.1063/1.1565325}.
\bibitem{Corezzi06}	Corezzi S., Comez L., Monaco G., Verbeni R., Fioretto D., Phys. Rev. Lett., 2006, \textbf{96}, 255702, \doi{10.1103/PhysRevLett.96.255702}.
\bibitem{Bencivenga} Bencivenga F., Inelastic Light and X-Ray Scattering from salol in the Supercooled, Glassy and Single Crystalline Phases, Universit\'a degli Studi di Perugia, Tesi di Laurea, 2003.
\bibitem{Zhang04} Zhang H.P., Brodin A., Barshilia H.C., Shen G.Q., Cummins H.Z., Pick R.M., Phys. Rev. E, 2004, \textbf{70}, 011502, \doi{10.1103/PhysRevE.70.011502}.
\bibitem{ISTS}Yang Y.,  Nelson K.A., J. Chem. Phys., 1995, \textbf{103}, 7732--7739, \doi{10.1063/1.470294}.
\bibitem{Maciovecchio98} Masciovecchio C., Monaco G., Ruocco G., Sette F., Cunsolo A., Krisch M., Mermet A., Soltwisch M., Verbeni~R., Phys. Rev. Lett., 1998, \textbf{80}, 544--547,
\doi{10.1103/PhysRevLett.80.544}.
\bibitem{Hansen98}  Hansen C., Stickel F., Richert R.,  Fischer E.W., J. Chem. Phys., 1998, \textbf{108}, 6408, \doi{10.1063/1.476063}.
\bibitem{Eckstein00}  Eckstein E.,  Qian J.,  Hentschke R.,  Thurn-Albrecht T.,  Steffen W.,  Fischer E.W., J. Chem. Phys., 2000, \textbf{113}, 4751, \doi{10.1063/1.1288907}.
\bibitem{Ruocco99}	Ruocco G., Sette F., Di Leonardo R., Fioretto D., Krisch M., Lorenzen M., Masciovecchio C., Monaco G., Pignon F.,  Scopigno T., Phys. Rev. Lett., 1999, \textbf{83}, 5583--5586, \doi{10.1103/PhysRevLett.83.5583}.
\bibitem{Stickel95} Stickel F.,  Fischer E.W.,  Richert R., J. Chem. Phys., 1995, \textbf{102}, 6251, \doi{10.1063/1.469071}.
\bibitem{Baran10} Baran J., Davydova N.A., Phys. Rev. E, 2010, \textbf{81}, 031503, \doi{10.1103/PhysRevE.81.031503}.
\bibitem{Scopigno03}	Scopigno T., Ruocco G., Sette F., Monaco G.,  Science, 2003, \textbf{302}, 849--852, \doi{10.1126/science.1089446}.
\bibitem{Buchenau04} Buchenau U., Wischnewski A., Phys. Rev. B, 2004, \textbf{70}, 092201, \doi{10.1103/PhysRevB.70.092201}.
\end{thebibliography}
\end{document}